\PassOptionsToPackage{table}{xcolor} %
\documentclass[sigconf]{acmart}
\usepackage[table]{xcolor}
\usepackage{amsmath}
\usepackage{tabularx}
\usepackage[utf8]{inputenc}
\usepackage{pifont}
\usepackage{tabularx}
\usepackage{booktabs}
\usepackage{xspace}
\usepackage{graphicx}
\usepackage{tikz}
\usetikzlibrary{fit,positioning,calc,shapes,backgrounds}
\usepackage{pgfplots}
\pgfplotsset{compat=1.14}
\usepgfplotslibrary{fillbetween}
\usepackage{epigraph}
\usepackage[caption=false]{subfig}
\usepackage{caption}
\usepackage{enumitem}
\usepackage[framemethod=tikz]{mdframed}
\usepackage{todonotes}
\usepackage{hyphenat} %
\usepackage{xfrac}
\usepackage{balance}

\DeclareMathOperator{\tr}{tr}

\newcommand\capi[1]{\ifnum\ifhmode\spacefactor\else2000\fi>1000 \uppercase{#1}\else#1\fi} %

\newcommand{\spara}[1]{\smallskip\noindent{\bf #1}}
\newcommand{\mpara}[1]{\medskip\noindent{\bf #1}}
\newcommand{\para}[1]{\noindent{\bf #1}}

\definecolor{cycle1}{RGB}{228, 26, 28}
\definecolor{cycle2}{RGB}{55, 126, 184}
\definecolor{cycle3}{RGB}{77, 175, 74}
\definecolor{cycle4}{RGB}{152, 78, 163}
\definecolor{cycle5}{RGB}{255, 127, 0}
\definecolor{cycle6}{RGB}{153, 153, 153}%
\definecolor{cycle7}{RGB}{166, 86, 40}
\definecolor{cycle8}{RGB}{247, 129, 191}

\newcommand*{\NPhard}{$\mathbf{NP}$-hard\xspace}
\newcommand*{\NPcomplete}{$\mathbf{NP}$-complete\xspace}

\newcommand*{\APXhard}{$\mathbf{APX}$-hard\xspace}
\newcommand*{\bigO}{\mathcal{O}}

\newcommand{\cmark}{\textcolor{cycle3}{\ding{52}}} %
\newcommand{\xmark}{\textcolor{cycle1}{\ding{56}}}

\newcommand{\thiswork}{\textsc{NetLSD}\xspace}
\newcommand{\ged}{\textsc{GED}\xspace}
\newcommand{\netsimile}{\textsc{NetSimile}\xspace}
\newcommand{\deltacon}{\textsc{DeltaCon}\xspace}
\newcommand{\mlgkernel}{\textsc{MLG}\xspace}
\newcommand{\fgsd}{\textsc{FGSD}\xspace}
\newcommand{\ftd}{\textsc{FTD}\xspace}

\newcommand*{\graphpair}[2]{(#1, #2)}

\newcommand{\reals}{\mathbb{R}}
\newcommand{\naturals}{\mathbb{N}}

\newcommand{\vertices}{V}

\newcommand{\edges}{E}
\newcommand{\graphcol}{\mathcal{G}}
\newcommand{\graph}{G}
\newcommand{\graphexp}[1]{\graph_{#1} = \graphpair{\vertices_{#1}}{\edges_{#1}}}
\newcommand{\sig}{\sigma}

\newcommand{\dist}{d}
\newcommand{\diam}{\mathbf{D}}

\newcommand{\mann}{\mathcal{M}}
\newcommand{\adj}{A}
\newcommand{\lapl}{\mathcal{L}}
\newcommand{\hk}{H_t}

\newcommand{\pk}[1]{{{#1}}} %
\newcommand{\ant}[1]{{{#1}}} %

\newcommand{\win}{\cellcolor{cycle3!30}}
\newcommand{\hide}[1]{} %

\newtheorem{property}{Property}
\newmdtheoremenv [%
	roundcorner=2pt,
    skipabove=6pt,
	linewidth=0.5pt,
	innertopmargin = 0pt,
    innerleftmargin = 5pt,
    innerrightmargin = 5pt,
	backgroundcolor = cycle2!10,%
] {remark}{Remark}

\copyrightyear{2018}
\acmYear{2018}
\setcopyright{acmlicensed}
\acmConference[KDD '18]{The 24th ACM SIGKDD International Conference on Knowledge Discovery \& Data Mining}{August 19--23, 2018}{London, United Kingdom}
\acmBooktitle{KDD '18: The 24th ACM SIGKDD International Conference on Knowledge Discovery \& Data Mining, August 19--23, 2018, London, United Kingdom}
\acmPrice{15.00}
\acmDOI{10.1145/3219819.3219991}
\acmISBN{978-1-4503-5552-0/18/08}

\title{Zen and the Art of Graph Comparison: A Spectral Approach} %
\title{NetLSD: Flexible Graph Comparison} %
\title{NetLSD: Laplacian Spectral Distance for Large Graph Collections} %
\title{NetLSD: Graph Vibrations as a Flexible Graph Similarity Measure} %
\title{NetLSD: Compare the Vibrations of Graphs} %
\title{The NetLSD Graph Similarity Measure:\\ Localized Vibrations on Various Scales of a Network} %
\title{NetLSD: Graph Comparison via Multi-Scale Localized Vibrations} %
\title{Expressive Graph Comparison via Multi-Scale Representations}
\title{Hearing the shape of the graph with NetLSD} %
\title{Comparing Graphs with NetLSD. How does that sound?} %
\title{NetLSD: Hearing a Graph's Multi-scale Structure} %
\title{NetLSD: Hearing the Shape of a Graph} %

\begin{document}
\author{Anton Tsitsulin}
\affiliation{%
  \institution{Hasso Plattner Institute}
}
\email{anton.tsitsulin@hpi.de}
\author{Davide Mottin}
\affiliation{%
  \institution{Hasso Plattner Institute}
}
  \email{davide.mottin@hpi.de}
\author{Panagiotis Karras}
\affiliation{%
  \institution{Aarhus University}
}
  \email{panos@cs.au.dk}
\author{Alex Bronstein}
\affiliation{%
  \institution{Technion}
}
  \email{bron@cs.technion.ac.il}
\author{Emmanuel M\"uller}
\affiliation{%
  \institution{Hasso Plattner Institute}
}
  \email{emmanuel.mueller@hpi.de}
\begin{abstract} %
\pk{Comparison among graphs is ubiquitous in graph analytics.} However, it is a hard task in terms of the expressiveness of the employed similarity measure and the efficiency of its computation.
Ideally, graph comparison should be invariant to the order of nodes and the sizes of compared graphs, adaptive to the scale of graph patterns, and scalable. Unfortunately, these properties have not been addressed together.
\pk{Graph comparisons still rely on direct approaches, graph kernels, or representation-based methods, which are all inefficient and impractical for large graph collections.}

In this paper, we propose the Network Laplacian Spectral Descriptor (NetLSD): the first, to our knowledge, permutation- and size-invariant, scale-adaptive, and efficiently computable graph representation method that allows for straightforward comparisons of large graphs.
NetLSD extracts a compact signature that inherits the formal properties of the Laplacian spectrum, specifically its heat \pk{or wave} kernel; thus, it {\em hears the shape of a graph\/}.
Our evaluation on a variety of real-world graphs demonstrates that it outperforms previous works in both expressiveness and efficiency.
\end{abstract} 
\maketitle

\section{Introduction}\label{sec:introduction}

Graphs are widely used for modeling complex structures such as biological networks, chemical compounds, social interactions, and knowledge bases. Such applications require the means to efficiently and meaningfully compare one graph to others.
Arguably, an ideal means for graph comparison should fulfill the following desiderata:
First, it should be indifferent to the order in which nodes are presented; we call this property {\bf permutation-invariance}.
Second, it would enable graph comparisons both at a \emph{local} level (expressing, e.g., atomic bond differences among chemical compounds) and at the \emph{global} or community level (capturing, e.g., the different topologies of social networks); we call this facility {\bf scale\hyp{}adaptivity}.
Third, it would detect structural similarity regardless of network magnitude (discerning, e.g., the similarity of two criminal networks of different size); we call this aptitude {\bf size\hyp{}invariance}.
Unfortunately, no existing means for graph comparison satisfies all three of these requirements.
Apart from these qualitative requirements, a viable means for graph comparison should be \emph{efficiently computable}.
Graph analytics tasks often require pairwise graph comparisons within a large collection of graphs, hence should be ideally done in \emph{constant time}, after preprocessing.
Unfortunately, existing methods fare even worse in this respect. A popular distance measure among graphs, \emph{graph edit distance} (\ged{})~\cite{sanfeliu1983}, defined as the minimum number of edit operations required to turn one graph into another, is \NPhard and \APXhard to compute~\cite{lin94}; intensive research efforts in \ged{}-based graph comparison~\cite{zeng09, zheng15, yuan15, gouda16, liang17} have not escaped this reality.
Similarly, \emph{graph kernel methods}~\cite{gartner2003, borgwardt2005, shervashidze2011, yanardag2015, nikolentzos2017, kondor2016} lack an explicit graph representation and they do not scale well either.
In this paper, we develop a permutation- and size-invariant, scale\hyp{}adaptive, and scalable method for graph comparison. We propose an expressive \emph{graph representation}, \thiswork, grounded on spectral graph theory (Sec.~\ref{sec:solution}), that allows for \emph{constant-time} similarity computations at several scales; \thiswork extracts compact graph signatures \pk{based on the heat or wave kernel of the Laplacian}, which inherit the formal properties of the Laplacian spectrum.

\begin{figure}[ht!]
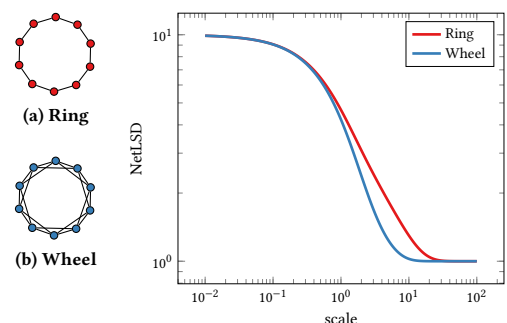

\vspace{-3mm}
\begin{minipage}[t]{\columnwidth}
    \centering
    \begin{minipage}[t]{0.13\columnwidth}
        \subfloat[Ring]{\resizebox{\linewidth}{!}{%
 %
\label{sfig:hksplot}}}
    \end{minipage}
\end{minipage}
\vspace{-4mm}
\caption{Comparison of similar graphs on multiple scales.}\label{fig:firstpage}
\vspace{-5mm}
\end{figure}

Figure~\ref{fig:firstpage} shows an example: at a small scale, a ring and a wheel are \pk{similar, as each node is connected to predecessor(s) and successor(s); at a large scale, they both appear as rings; at intermediate scales their local structures differ}.
\begin{table*}[ht]
\setlength{\aboverulesep}{0pt}
\setlength{\belowrulesep}{0pt}

\begin{center}
{
\small
\newcolumntype{C}{>{\centering\arraybackslash}X}
\begin{tabularx}{\textwidth}{Xp{2.75cm}CCCCC} %
\multicolumn{1}{C}{} & \multicolumn{1}{C}{} & \multicolumn{3}{c}{\textbf{Properties}} & \multicolumn{2}{c}{\textbf{Complexity}} \\
\cmidrule(lr){3-5}\cmidrule(lr){6-7}
\bf{Method} & {\bf Type} & Perm.\hyp{}invariance & Scale\hyp{}adaptivity & Size\hyp{}Invariance & Precomputation & Comparison\\ 
\midrule
\ged{}~\cite{sanfeliu1983}     & direct & \cmark{} & \xmark{}    & \xmark{}     & $\bigO(1)$            & \NPcomplete{}\\ %
\deltacon{}~\cite{koutra2013}  & direct & \xmark{} & \xmark{}    & \xmark{}     & $\bigO(1)$            & $\bigO(m)$\\
\ftd~\cite{bento2018family}    & direct & \xmark   & \xmark      & \xmark		& $\bigO(1)$            &  $\bigO(n^2)$\\
\midrule{}
SP~\cite{borgwardt2005}        & kernel & \cmark{} & \xmark{}    & \xmark{}     & $\bigO(1)$            &  $\bigO(n^3)$\\
\mlgkernel{}~\cite{kondor2016} & kernel & \cmark{} & \cmark{}    & \xmark{}     & $\bigO(k m + k^2 n)$  & $\bigO(k^3)$\\
\midrule{}
\netsimile{}~\cite{berlingerio2013} & statistical representation & \cmark{} & \xmark{} & \cmark{} & $\bigO(n \log n)$ & $\bigO(1)$\\
\fgsd{}~\cite{verma2017} & statistical representation & \cmark{} & \xmark{} & \xmark{} & $\bigO(n^2)$ & $\bigO(1)$\\
\midrule
\rowcolor{cycle2!10}
\textbf{\thiswork{}} & spectral representation & \cmark{} & \cmark{} & \cmark{} & $\bigO(k m + k^2 n)$ & $\bigO(1)$\\
\end{tabularx}
}
\end{center}
\caption{Related work in terms of fulfilled (\cmark) and missing (\xmark) properties, complexity ($n$ nodes, $m$ edges, $k$ eigenvalues).}\label{tbl:relatedwork}
\vspace{-3mm}
\end{table*} 

\hide{
We also introduce \pk{approximations of these trace signatures} (Sec.~\ref{ssec:computation}) and a normalization scheme for the sake of size-invariance (Sec.~\ref{ssec:properties}).
Our thorough empirical evaluation with real data (Sec.~\ref{sec:experiments}) demonstrates that \thiswork outperforms previous works in expressiveness and efficiency.}
\section{Related Work}\label{sec:related-work}

\pk{We distinguish  methods for graph comparison into three categories:
\emph{direct}, %
\emph{kernel methods}, and %
\emph{statistical representations}. %
To the best of our knowledge, this is the first work employing a \emph{spectral representation} for graph comparison.}
\subsection{Direct Methods}

\emph{Graph edit distance} (\ged)~\cite{sanfeliu1983} is the minimal number of edit operations needed to transform one graph into another; unfortunately, its calculation is \NPhard~\cite{garey2002computers} and even hard to approximate (\APXhard)~\cite{lin94}, as it implies determining a \emph{correspondence} among the compared graphs' nodes, a computationally hard task to begin with~\cite{gartner2003, shervashidze2011, yanardag2015, nikolentzos2017, kondor2016}; some application-specific techniques (e.g., anomaly detection in time evolving graphs) impose assumptions on that correspondence to alleviate the burden~\cite{koutra2013, papadimitriou2010, bento2018family}; while \ged admits heuristic approximation~\cite{riesen2009approximate, fischer2015approximation} and indexing schemes~\cite{zeng09, zhu2012finding,liang17}, it is not applicable for generic comparison among large graph collections. %
Besides, \ged{} treats all edit operations as equal, without discerning the extent to which they may alter the graph topology.
Thus, even if the computational obstacle were surmounted, \ged would still be unsatisfactory as a measure for multi-scale and multi-size graph comparisons.
In an effort to overcome \ged's shortcomings, past works have resorted to more flexible distance definitions, based on vertex and path similarities~\cite{papadimitriou2010} or propagation models~\cite{koutra2013}.
While such measures are less sensitive to local changes than \ged, they usually require node to be aligned in advance, hence are limited to specific applications.
An attempt to abandon this requirement relaxes the permutation matrix required to define a full node correspondence and considers a family of tractable distances (FTD)~\cite{bento2018family}.
This relaxed matrix preserves local properties and is easy to compute, %
yet the ensuing measure is still sensitive to permutations.
\subsection{Kernel Methods}

\emph{Graph kernels}~\cite{gartner2003, borgwardt2005, shervashidze2011, yanardag2015, nikolentzos2017, kondor2016} are similarity functions among graphs, which typically perform an implicit transformation of graph structure to compare two graphs (e.g.\ Shortest-path (SP) kernel~\cite{shervashidze2011}).
While each kernel function has its own valuable properties, to our knowledge, no extent graph kernel achieves both scale\hyp{}adaptive and size-invariant graph comparison.
Besides, kernels require expensive on-demand computations at comparison time, hence are inapplicable for large-scale graph comparisons.
Recently, Kondor and Pan proposed the Multi-scale Laplacian Graph kernel (MLG)~\cite{kondor2016}, which achieves scale\hyp{}adaptivity by exploiting the propagation of information in the graph and summing the information at each iteration.
However, it also raises a computational overhead cubic in Laplacian matrix eigenvalues.
\subsection{\pk{Statistical Representations}}

Representation\hyp{}based methods generate an one-off graph signature vector, based on statistical properties, and use it in subsequent inter-graph comparisons. \pk{Preliminary works in this area~\cite{bronstein2011a, berlingerio2013, bonner2016} handcraft features by aggregating local graph properties such as a node's and its neighbors' degrees. Such representations are easy to compute, yet focus on local characteristics, and are oblivious to global features.}
\pk{A more advanced approach, the Family of Spectral Distances (FSGD),~\cite{verma2017} produces a high-dimensional sparse representation as a histogram on the dense biharmonic graph kernel; however, FGSD does not capture graph features at different scales of resolution or graph sizes, and is also inapplicable to reasonably large graphs, due to its quadratic time complexity.}
\subsection{\pk{Spectral Representations}}

\emph{Spectral graph theory} is effective in the comparison of 3D objects~\cite{osada2002, gal2007, bronstein2011a}.
Bronstein~et~al.~\cite{bronstein2006} proposed computing pairwise shape similarities by finding the minimum-distortion embedding of one shape into the other. This procedure
is related to the computation of the Gromov-Hausdorff distance~\cite{memoli2005theoretical}, which is suitable for shape comparison, yet hard to compute.
Most relevant works consider filtering functions corresponding to known diffusion models, including heat~\cite{sun2009,bronstein2011b}, wave~\cite{aubry2011} and commute time~\cite{bronstein2011b} distances.
While it is known that hitting and commute times degenerate to a function of node degree on large graphs~\cite{von2010}, it is unknown whether other filters designed for three\hyp{}dimensional manifolds maintain their expressiveness on high-dimensional graphs.

While 3D objects have a precise low-dimensional shape, graphs have no rigid form. Yet we can consider a graph as a geometrical object.
G\"{u}nthard and Primas first asked to what extent a graph (in general, a manifold) may be determined by its spectrum~\cite{gunthard1956}; Kac expressed the same question eloquently: ``Can one {\em hear the shape} of a drum?''~\cite{kac1966}.
Since then, investigations have shown that isospectral graphs tend to be isometric~\cite{wilson2008}, and some graphs are determined by their spectrum~\cite{vandam2003}.
Thus, spectral graph theory forms a solid ground for graph comparison. The Laplacian spectrum is used in graph mining~\cite{shi2000normalized, akoglu2010}, yet, to our knowledge, no previous work has used it as a means for graph comparison. %

\section{Problem Statement}\label{sec:problem-statement}

An undirected graph is a pair $\graphexp{}$, where $\vertices = (v_1, \ldots, v_n), n = |\vertices|$ is the set of vertices and $\edges\subseteq (\vertices \times \vertices)$ the set of edges. %
We assume the graph is unweighted even though our method readily applies to the weighted case.
A \emph{representation} is a function $\sig : \graphcol \rightarrow \reals^\naturals$ from any graph $\graph$ in a collection of graphs $\graphcol$ to an infinitely dimensional real-vector; the element $j$ of the representation is denoted as $\sig_j(\graph)$.
A \emph{representation-based distance} is a function $\dist^\sig: \reals^\naturals \times \reals^\naturals \rightarrow \reals_0^+$ on the representations of two graphs $\graph_1, \graph_2 \in \graphcol$ that returns a positive real number. We aim to devise a constant-time \emph{representation-based distance} among any pair of graphs $\graph_1, \graph_2$. %

\subsection{Expressive Graph Comparison}\label{ssec:formal-requirements}

Our distance should support data mining tasks, such as clustering, nearest neighbor classification, distance-based anomaly detection.
Therefore, we require our distance to be a \emph{pseudometric}; namely, it should fulfill the following properties. 

\begin{itemize}
    \item \emph{Symmetry}, for any $\graph_1, \graph_2 \in \graphcol$:\\ $\dist^\sig(\sig(\graph_1),\sig(\graph_2)) = \dist^\sig(\sig(\graph_2),\sig(\graph_1))$ %
    \item \emph{Triangle inequality}, for any $\graph_1, \graph_2, \graph_3 \in \graphcol$:\\ $\dist^\sig(\sig(\graph_1),\sig(\graph_3)) \le  \dist^\sig(\sig(\graph_1),\sig(\graph_2)) + \dist^\sig(\sig(\graph_2),\sig(\graph_3))$ %
\end{itemize}

These properties characterize a large family of distances, yet do not reflect their expressiveness.
We require \textit{expressive distances} to be permutation-invariant, scale\hyp{}adaptive, and size-invariant. 

\spara{Permutation-invariance} implies that if two graphs' structure are the same (i.e., if the two graphs are isomorphic) the distance of their representations is zero.
A graph $\graphexp{1}$ is isomorphic to another graph $\graphexp{2}$, or $\graph_1 \simeq \graph_2$, if there exists a bijective function $\mu : \vertices_1 \rightarrow \vertices_2$ such that $(\mu(u),\mu(v)){\in}\edges_2$ for each $(u,v){\in}\edges_1$.

\begin{property}[Permutation-invariance] A distance $\dist^\sig$ on representation $\sig$ is permutation-invariant iff: 
\[\forall \graph_1,\graph_2,~\graph_1 \simeq \graph_2 \Rightarrow \dist^\sig(\sig(\graph_1), \sig(\graph_2)) = 0\]
\end{property}

\para{Scale\hyp{}adaptivity} implies that a representation accounts for both local (edge and node) and global (community) graph features.
A global feature cannot be captured by \emph{any} combination of features on nodes at distance $r < \diam(\graph) - 1$, where $\diam(\graph)$ is the diameter (longest shortest-path length) of $\graph$. %
Let the set of all subgraphs of $\graph$ be $\xi(G)=\{g \sqsubset G: \diam(g)<\diam(\graph)\}$.
We define scale\hyp{}adaptivity as the property of a representation $\sig$ having at least one local feature (i.e., derived only from information encoded in subgraphs $\xi(G)$), and at least one global feature (i.e., derived by strictly more than the information encoded in any $\xi(G)$).
Using local features only, a similarity measure would deem two graphs sharing local patterns to have near-zero distance although their global properties (such a page-rank features) may differ, and, in reverse, relying on global features only would miss local structures (such as edge distributions).
We aim for a representation adaptive to both local and global structures on demand.

\begin{property}[Scale\hyp{}adaptivity]
A representation $\sig$ is scale\hyp{}adaptive iff it contains both local features $\sig_i$ and global features $\sig_j$: 
\begin{itemize}%
	\item Local Feature: $\forall \graph \: \exists f(\cdot):\: \sig_i=f(\xi(G))$
	\item Global Feature: $\forall \graph \: \not\exists f(\cdot):\: \sig_j=f(\xi(G))$
\end{itemize}
\end{property}

\para{Size-invariance} is the capacity to discern that two graphs represent the same phenomenon at a different magnitudes (e.g., two criminal circles of similar structures but different sizes should have near-zero distance).
We can think of a graph as a representation of a metric space (a \emph{manifold} in particular) with a small intrinsic dimension.
We would then like to abstract away the particular way of sampling that space. Size-invariance postulates that if two graphs originate from the sampling of the same domain $\mann$, they should be deemed similar.

\begin{property}[Size-invariance]
A size-invariant distance $\dist^\sig$ on representation $\sig$ fulfills:
\[\forall \mann: \graph_1, \graph_2 \mbox{ sampled from } \mann \Rightarrow
\dist^\sig(\sig(\graph_1),\sig(\graph_2)) = 0\] %
\end{property}

An expressive means of graph comparison should employ a representation fulfilling the above properties.
\thiswork, presented in the sequel, is \pk{a graph representation that allows for easy and expressive graph comparison.}
\section{NetLSD: Network Laplacian Spectral Descriptor}\label{sec:solution} 

Defining a representation fulfilling the requirements of permutation\hyp{}, scale-, size- invariance, and efficiency is tough; in general, structures are hard to compare.
Thus, we transfer the problem to the spectral domain. 
A useful metaphor is \pk{that} of heating the graph's nodes and observing the heat diffusion as time passes.
Another useful metaphor is that of a system of masses corresponding to the graph's nodes and springs corresponding to its edges. The propagation of mechanical waves through the graph is another way to capture its structural invariants.
\pk{In both cases, the overall process describes the graph in a permutation\hyp{}invariant manner, and embodies more global information as time elapses. Our representation employs a \emph{trace signature} encoding such a {\em heat diffusion\/} or {\em wave propagation\/} process over time. We compare two graphs via the $L_2$ distance among trace signatures sampled at selected time scales.}

\subsection{Spectra as representations}\label{ssec:representation}

The \emph{adjacency matrix} of a graph $\graph$ is a $n\times n$ matrix $\adj$ having $\adj_{ij} \!=\! 1$ if $(i,j) \in \edges$ and $\adj_{ij}\! =\! 0$ otherwise.
A graph's \emph{normalized Laplacian} is the matrix $\lapl\! =\!  I\! -\! D^{-\frac{1}{2}}AD^{-\frac{1}{2}}$, where $D$ is the diagonal matrix with the degree of node $i$ as entry $D_{ii}$, i.e, $D_{ii} = \sum_{j = 1}^n A_{ij}$.
Since the Laplacian is a symmetric matrix, its eigenvectors $\phi_1, \ldots, \phi_n$, are real and orthogonal to each other.
Thus, it is factorized as $\lapl = \Phi\Lambda\Phi^\top$, where $\Lambda$ is a diagonal matrix on the sorted eigenvalues $\lambda_1 \le \ldots \le \lambda_n$ of which $\phi_1, \ldots, \phi_n$ are the corresponding eigenvectors, and $\Phi$ is an orthogonal matrix obtained by stacking the eigenvectors in columns $\Phi = [\phi_1 \phi_2 \ldots \phi_n]$.

The set of eigenvalues $\{\lambda_1, \ldots, \lambda_n\}$ is called the \emph{spectrum} of a graph.
The Normalized Laplacian, as opposed to the unnormalized version $L\! =\! D \!-\! A$, has a bounded spectrum, $0 \le \lambda_i \le 2$. 

Belkin~and~Niyogi~\cite{belkin2007} showed that eigenvectors of the normalized Laplacian of a point cloud graph converge to the eigenfunction of the Laplace-Beltrami operator~\cite{berger2012panoramic} on the underlying Riemannian manifold.
In general, the normalized Laplacian has more attractive theoretical properties than its unnormalized counterparts~\cite{vonluxburg2008}.

The Laplacian spectrum encodes important graph properties, such as the normalized cut size~\cite{shi2000normalized} used in spectral clustering. \pk{Likewise, the normalized Laplacian spectrum can determine whether a graph is bipartite, but not the number of its edges~\cite{chung1997}. Rather than using the Laplacian spectrum per se, we consider an associated heat diffusion process on the graph, to obtain a more expressive representation, in a manner reminiscent of random walk models~\cite{chung2007}}.

The \emph{heat equation} associated with the Laplacian is 
\begin{equation}\label{eq:heat-diffeq}
    \frac{\partial u_t}{\partial t} = -\mathcal{L}u_t,
\end{equation}
\noindent where $u_t$ are scalar values on vertices representing the heat of each vertex at time $t$.
The solution to the heat equation provides the heat at each vertex at time $t$, when the initial heat $u_0$ is initialized with a fixed value on one of the vertices.
Its closed-form solution is given by the $n\times n$ \emph{heat kernel} matrix:
\begin{equation}\label{eq:heat-solution}
    \hk = e^{-t\mathcal{L}} = \sum_{j=1}^n e^{-t \lambda_j} \phi_j \phi_j^\top,
\end{equation}
where ${(\hk)}_{ij}$ represents the amount of heat transferred from vertex $v_i$ to vertex $v_j$ at time $t$. 
We can also compute the heat kernel matrix directly by exponentiating the Laplacian eigenspectrum~\cite{chung1997}:
\begin{equation}\label{eq:heat-solution-eigen}
    H_t = \Phi{}e^{-t\Lambda}\Phi^\top
\end{equation}
As the heat kernel involves pairs of nodes, it is not directly usable to compare graphs.
We rather consider the \emph{heat trace} at time $t$:
\begin{equation}\label{eq:heat-trace}
    h_t = \tr(\hk) = \sum_{j}{e^{-t\lambda_j}}
\end{equation}

Then our \thiswork representation consists of a {\bf heat trace signature} of graph $\graph$, i.e., a collection of heat traces at different \emph{time scales}, $h(\graph) = {\{ h_t \}}_{t>0}$.

\mpara{Alternative signatures.} 
The heat kernel can be viewed as a family of \emph{low-pass filters}, $F(\lambda) = e^{-\lambda t}$, parametrized by the scale parameter $t$, hence the heat trace signature contains low frequency (i.e., large-scale) information at every scale.
Other kernels deemphasize the influence of low frequencies. For example, the \emph{wave equation},
\begin{equation}\label{eq:wave-diffeq}
    \frac{\partial^2 u_t}{\partial t^2} = -\mathcal{L}u_t,
\end{equation}
which describes the amplitude $u_t$ of a wave propagating in a medium, has, in its turn, a solution given by the \emph{wave kernel}:
\begin{equation}\label{eq:wave-solution}
    W_t = e^{-i t\mathcal{L}} = \sum_{j=1}^n e^{-it \lambda_j} \phi_j \phi_j^\top
\end{equation}
(note the complex exponential) and a corresponding {\bf wave trace signature} with $t \in [0,2\pi)$:
\begin{equation}\label{eq:wave-trace}
    w_t = \tr(W_t) = \sum_{j}{e^{-it\lambda_j}}
\end{equation}
\subsection{Scaling to large graphs}\label{ssec:computation}

\pk{The full} eigendecomposition of the Laplacian \(\mathcal{L} = \Phi{}\Lambda\Phi^\top{}\) takes \(\bigO(n^3)\) time and \(\Theta(n^2)\) memory.
This allows to compute signatures of graphs with over a thousand nodes in less than a second on commodity hardware, yet renders direct computation impossible for larger graphs.
Thus, we need to approximate heat trace signatures.
We propose two different methods to that end. 

\pk{Our first proposal is to use a {\bf Taylor expansion}; while this mathematical tool provides a rather dubious approximation of a matrix exponential, as its convergence rate depends on the largest eigenvalue~\cite{moler2003}, it is useful on \emph{small time scales} \(t\),
and allows for an inexpensive computation of its first two terms,}
\begin{equation}\label{eq:hks-taylor}
h_t = \sum_{k=0}^{\infty}{\frac{\tr({(-t\mathcal{L})}^k)}{k!}}
    \approx n - t\, \tr(\mathcal{L}) + \frac{t^2}{2}\tr(\mathcal{L}^2)
\end{equation} %

\noindent \pk{These first two terms are easily computed, even for very large graphs, as $\tr(\mathcal{L})=n$ and \(\tr\left(\mathcal{L}^2\right) = \sum_{ij}{{\mathcal{L}_{ij}}^2}\) since \(\mathcal{L}\) is self-adjoint.}
This way, we can compare two graphs locally in \(\bigO(m)\).

\begin{figure}[ht!]
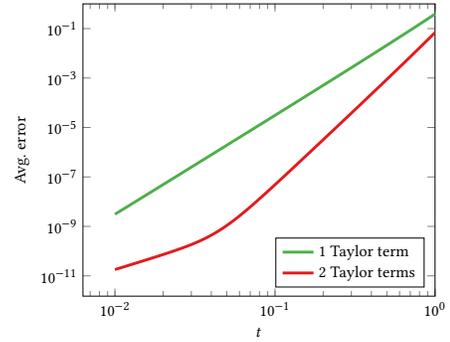

\centering
\resizebox{0.7\columnwidth}{!}{%
 %
}%
\caption{Relative approximation error of normalized $h_t$ for Erd{\H{o}}s-R{\'e}nyi random graphs, varying time scale $t$.}\label{fig:approximation-accuracy-taylor}
\end{figure}

\begin{figure}[ht!]
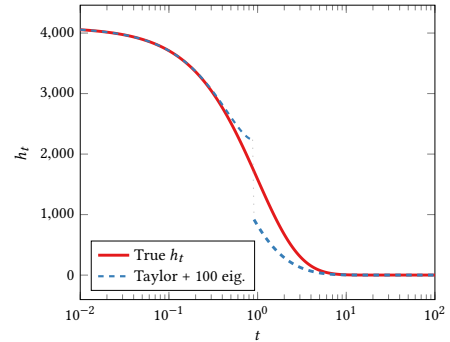

\centering
\resizebox{0.7\columnwidth}{!}{%
 %
}%
\caption{{}Heat trace approximation by two Taylor terms and 100 eigenvalues on a random blockmodel~\cite{karrer2011} graph.}\label{fig:approximation-accuracy-frankenstein}
\end{figure}

\pk{Figure~\ref{fig:approximation-accuracy-taylor} depicts the error in approximating the normalized heat trace by a Taylor expansion for random graphs of varying sizes; this error is independent of graph size, and stays low until time scale 1.}
At \emph{large\ant{r} time scales}, the influence of high frequencies (i.e., the higher part of the spectrum) on the heat trace decreases. Thus, \pk{one can} approximate the heat trace signature using the lower part of the eigenspectrum, as in shape analysis~\cite{sun2009, vaxman2010, bronstein2011b, litman14}. \pk{Thus}, we \pk{may} apply the low-order Taylor expansion for \emph{small $t$} and the truncated spectrum approximation for \emph{large $t$}. However, this approach misses out on the medium scale, as Figure~\ref{fig:approximation-accuracy-frankenstein} illustrates.
Besides, this technique does not lend itself to comparing graphs with different numbers of computed eigenvalues, since the spectrum discretization ratio is not normalized across the networks.

\begin{figure*}[ht!]
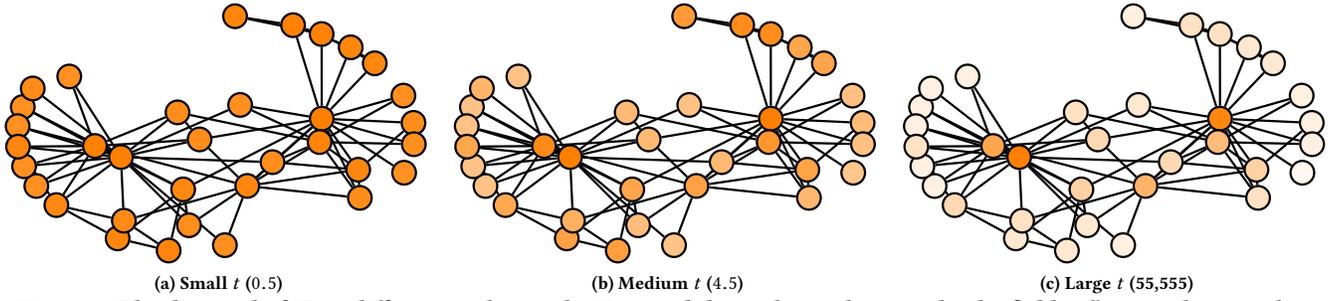

\centering
\subfloat[Small \(t\) ($0.5$)]{
\resizebox{0.32\textwidth}{!}{%
%
 %
}\label{sfig:karate-large}
}%
\vspace{-4mm}
\caption{The diagonal of $H_t$ at different scales on the Karate club graph; at a large scale, the field reflects node centrality.}\label{fig:karate-club}
\vspace{-2mm}
\end{figure*} %
\begin{figure}[h]
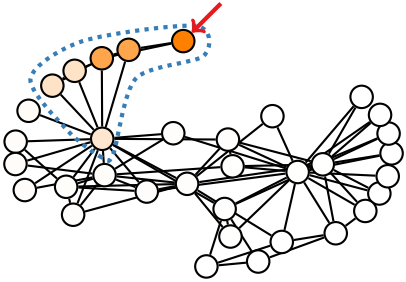

\vspace{-2mm}
\centering
%
%
\vspace{-2mm}
\caption{Visualization of the diagonal in the heat kernel matrix $H_t$ for the pointed vertex at scale $t\!=\!0.3$
}\label{fig:karate-club-locality}%
\vspace{-4mm}
\end{figure} 
\pk{We conclude that the Taylor expansion is useful on very large graphs, on which eigendecomposition is prohibitive. For manageable graph sizes, we adopt a more accurate strategy~\cite{moler2003}
based on approximating} the \emph{eigenvalue growth rate}, as in~\cite{vaxman2010}:
we compute $k$ eigenvalues on \emph{both} ends of the spectrum, and interpolate a \emph{linear growth} of the \pk{interloping} eigenvalues.
This strategy assumes that on the medium scale the manifold defining the graph is two-dimensional, as Weyl's law of asymptotic eigenvalue growth suggests~\cite{weyl1911}.
Since we only need to compute extreme eigenvalues,
we use the block Krylov-Schur implementation in SLEPc~\cite{hernandez2005}.
Graph Laplacians always have a zero eigenvalue with eigenspace dimension equal to the number of connected components; \pk{we deflate the search space for the eigensolver thanks to this property}. In our experimental study, we employ this {\em interpolation technique;} \pk{In Section~\ref{ssec:approximation-quality}, we evaluate its approximation quality.}
\subsection{Properties of the heat trace}\label{ssec:properties}

Here, we discuss how the heat trace signatures achieves our target representation properties.

\mpara{Permutation-invariance.} The permutation invariance of $h(\graph)$ follows from the properties of the spectrum: isomorphic graphs are isospectral, hence their respective heat trace signatures are equal. 

\mpara{Scale\hyp{}adaptivity.} The heat kernel can be seen as continuous-time random walk propagation, and its diagonal (sometimes referred to as the autodiffusivity function or the heat kernel signature) can be seen as a continuous-time PageRank~\cite{chung2007}.
Figure~\ref{fig:karate-club} shows the heat kernel signature with small (a), medium (b), and large (c) $t$; with large $t$, the heat tends to focus on central nodes.

As \(t\) approaches zero, the Taylor expansion yields \(H_t \simeq I - \mathcal{L}t\), meaning the heat kernel depicts \emph{local connectivity}. On the other hand, for large \(t\),
$H_t \simeq I - e^{-\lambda_2t}\phi_2{\phi_2}^\top$,
where $\phi_2$ is the \emph{Fiedler vector} used in spectral graph clustering~\cite{shi2000normalized}, as it encodes \emph{global connectivity}.
Thus, the heat kernel localizes around its diagonal, and the degree of localization depends on the scale $t$; it can thereby be tuned to produce both local and global features.

Figure~\ref{fig:karate-club-locality} illustrates heat kernel locality, focusing on a single row of $H_{t}$ corresponding to the node marked with a red arrow; the kernel is localized in the region marked with a dotted line.

\mpara{Size-invariance.} While the normalized Laplacian alleviates the problem of different edge densities~\cite{chung1997}, the Taylor expansion in Equation~\ref{eq:hks-taylor} manifests that \(h(G)\) contains information about \pk{the} number of nodes.
Fortunately, we can employ \emph{neutral} graphs for normalization, namely the empty and the complete graph with $n$ nodes.
Eigenvalues of the normalized Laplacian of an empty graph \(\bar{K}_n\) of size \(n\) are all zero; for a complete graph \(K_n\), they are given by a vector of one zero and \(n\!-\!1\) ones.
Thus, the heat traces of these graphs are analytically computed as:
\begin{align*}
    h_t(\bar{K}_n) &= \frac{1}{n} \\
    h_t(K_n) &= 1 + (n-1) e^{-t}.
\end{align*}
\noindent and their wave traces are:
\begin{align*}
    w_t(\bar{K}_n) &= \frac{1}{n} \\
    w_t(K_n) &= 1 + (n-1)\cos(t). 
\end{align*}
Either option can be used to normalize the heat or wave trace signatures.
Such normalization can be interpreted as a modification of the corresponding diffusivity tensor (for the heat trace) or the elasticity tensor (for the wave trace).
We provide more details in Section~\ref{sec:experiments}.
\begin{figure*}[!t]
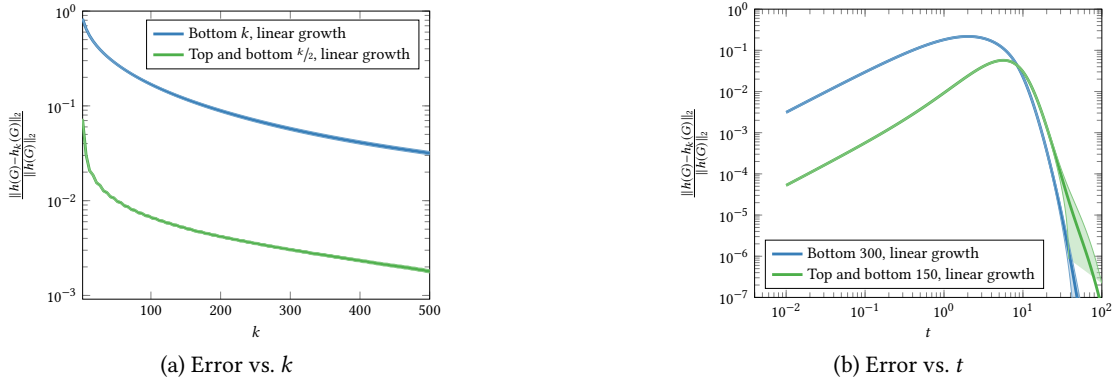

\vspace{-4mm}
\centering

\vspace{-3mm}
\caption{Relative error in spectrum computation of $h(G)$, averaged across 2085 graphs.}\label{fig:approximation-accuracy-diff}
\vspace{-4mm}
\end{figure*} %
\subsection{Connection to computational geometry}\label{ssec:theory}

M{\'e}moli~\cite{memoli2011} suggests a spectral definition of the Gromov\hyp{}Wasserstein distance between Riemannian manifolds based on matching the heat kernels at all scales.
The cost of matching a pair of points $(x,x')$ on manifold $\mathcal{M}$ to a pair of points $(y,y')$ on manifold $\mathcal{N}$ at scale $t$ is given by
$$
\Gamma(x,y,x',y',t) = | H^\mathcal{M}_t(x,x') - H^\mathcal{N}_t(y,y') |.
$$
The distance between the manifolds is then defined in terms of the infimal measure coupling 
$$
d(\mathcal{M},\mathcal{N}) = \inf_{\mu} \sup_{t>0} e^{-2(t+t^{-1})} \, \| \Gamma \|_{L^2(\mu \times \mu)},
$$
where the infimum is sought over all measures on $\mathcal{M} \times \mathcal{N}$ marginalizing to the standard measures on $\mathcal{M}$ and $\mathcal{N}$.
M{\'e}moli~\cite{memoli2011} shows that this distance can be lower bounded by
$$
d(\mathcal{M},\mathcal{N})  \ge \sup_{t>0} e^{-2(t+t^{-1})} \, | h^\mathcal{M}_t - h^\mathcal{N}_t |.
$$
In other words, the lower bound is the scaled $L_\infty$ distance between the heat trace signatures $h(\mathcal{M})$ and $h(\mathcal{N})$. 

We effectively adopt this result \emph{mutatis mutandis} to graphs, substituting the Laplace-Beltrami operator of the manifold with the normalized graph Laplacian; we then sample the heat trace signature at a finite number of scales, rendering it a versatile vector representation of a graph, as other works produce vector representations of graph vertices~\cite{ribeiro17, tsitsulin18}.
This lower bound implies that, if the distance between heat trace signatures is sufficiently large, the compared graphs cannot be similar.
Unlike the spectral Gromov-Wasserstein distance, distances between heat trace signatures are easily \emph{indexable}; thus, the derived bound allows for efficiently pruning of dissimilar graphs while working with the index alone, as with any lower-bounding scheme for high-dimensional search~\cite{tao09, kashyap11}.
\begin{table}[h!]
\vspace{-4mm}
\begin{center}
{
\newcolumntype{C}{>{\centering\arraybackslash}X}
\begin{tabularx}{\columnwidth}{p{1.75cm}CCCCC} %
& & & \multicolumn{3}{c}{vertices $|\vertices|$}\\
\cmidrule(lr){4-6}
\emph{dataset} & $|\graphcol|$ & $|Y|$ & min & avg & max \\ 
\midrule
MUTAG & 188 & 2 & 10 & 17.93 & 28 \\
PTC & 344 & 2 & 2 & 25.56 & 109 \\
PROTEINS & 1113 & 2 & 4 & 39.06 & 620 \\
NCI1 & 4110 & 2 & 3 & 29.87 & 111 \\
NCI109 & 4127 & 2 & 4 & 29.68 & 111 \\
ENZYMES & 600 & 6 & 2 & 32.63 & 126 \\
D\&D & 1178 & 2 & 30 & 284.3 & 5748 \\
\midrule
COLLAB & 5000 & 3 & 32 & 74.49 & 492 \\
IMDB-B & 1000 & 2 & 12 & 19.77 & 136 \\
IMDB-M & 1500 & 3 & 7 & 13.00 & 89 \\
REDDIT-S & 9543 & 2 & 100 & 337.8 & 999 \\
REDDIT-M & 2000 & 2 & 6 & 429.6 & 3782 \\
REDDIT-X & 2085 & --- & 1001 & 1447.2 & 5242 \\
REDDIT-L & 291 & 2 & 4445 & 97491 & 1632141 \\
\bottomrule
\end{tabularx}
}
\end{center}
\caption{Dataset properties.}\label{tbl:datasets}
\vspace{-10mm}
\end{table} 
\section{Experimental evaluation}\label{sec:experiments}

We evaluate\footnote{Source code and data are available at
\url{https://github.com/xgfs/NetLSD}} \thiswork{} against \netsimile~\cite{berlingerio2013} and \fgsd~\cite{verma2017}.
We run the experiments on a 20-core Intel Xeon CPU E5-2640v4, 3.20GHz machine with 256Gb RAM. Each method is assessed on the \emph{best parameters} through cross-validation. 
We require each method to complete within one day, or else an early termination is issued.
\ant{We used graph-tool~\cite{peixoto2014} for graph manipulation and synthetic graph generation.}

\mpara{Benchmarks.} We compare to the following comparison methods.

\spara{\netsimile{}~\cite{berlingerio2013}}: \pk{a representation using handcrafted features} obtained by aggregating statistics on nodes and edges (e.g., average degree, standard deviation of the degree of the neighbors). For each graph, the resulting representation has 35 dimensions. As recommended in~\cite{berlingerio2013}, we use the Canberra\footnote{\url{https://en.wikipedia.org/wiki/Canberra_distance}} distance for comparison. 

\spara{\fgsd{}~\cite{verma2017}}: a method that computes histograms on the biharmonic kernel of the graph. 
Such histograms typically bear large dimensionality (${\ge}50.000$) using the recommended bin-width $0.0001$.

\smallskip

\mpara{Parameter settings.}
Unless otherwise stated, we repeat each experiment $100$ times and report the average across all trials.
\thiswork is instantiated to both heat $h(\graph)$ and wave $w(\graph)$ trace signature described Section~\ref{ssec:representation}. 
In addition, we evaluate our normalized versions: the normalization with empty graph $h(G)/h(\bar{K})$, $w(G)/w(\bar{K})$ and the normalization with complete graph $h(G)/h(K)$, $w(G)/w(K)$.
To build \thiswork signature vectors, we need to sample a number of traces, i.e., values of $t$. After experimentation, we settled for $250$ values evenly spaced on the logarithmic scale in the range $[10^{-2}, 10^2]$; we attested this to be a good choice in terms of the quality\hyp{}size tradeoff, hence use these settings in all experiments\pk{, with both heat and wave signatures}. 
With small graphs, we employ the full eigendecomposition to produce the trace for each $t$ value. \pk{With the larger REDDIT-L graph}, we use 300 eigenvalues, 150 from each side of the eigenspectrum, by default, unless indicated otherwise; \pk{we validate these choices in Section~\ref{ssec:approximation-quality} (cf. Figure~\ref{fig:approximation-accuracy-diff}).}

\mpara{Datasets.}
We use six graph collections from bioinformatics~\cite{yanardag2015} and three social networks.
COLLAB, and IMDB-(B/M) graph collections are obtained sampling neighborhoods of nodes in a collaboration and movie network, respectively. 
REDDIT-S, REDDIT-M, and REDDIT-L are selected among small (at most $1000$ nodes), medium (at most $4000$ nodes) and large ($> 4000$ nodes) subreddits\footnote{https://dynamics.cs.washington.edu/data.html}, respectively. 
We report their main characteristics in Table~\ref{tbl:datasets}: number of graphs $|\graphcol|$, number of labels for classification $|Y|$, minimum, average, and maximum number of vertices. 

\begin{table*}[ht]
\begin{center}
{
\newcolumntype{C}{>{\centering\arraybackslash}X}
\begin{tabularx}{\textwidth}{p{1.5cm}CCCCCCCC} %
\emph{dataset}  & $h(G)$ & $h(G)/h(\bar{K})$ & $h(G)/h(K)$ & $w(G)$ & $w(G)/w(\bar{K})$ & $w(G)/w(K)$  & \fgsd{} & \netsimile{} \\ 
\midrule
MUTAG           & 76.03 & 79.12 & 78.22 & 78.18 & \win 79.72 & \win 79.38 & 77.79 & 77.11 \\
PTC             & 56.41 & 62.53 & 63.11 & 58.55 & \win 64.28 & 60.46 & 54.75 & 62.12 \\
PROTEINS        & 91.81 & \win 94.90 & \win 95.31 & 93.04 & 89.00 & 91.27 & 60.11 & 85.73 \\
NCI1            & 69.74 & \win 74.55 & 69.89 & 70.54 & \win 74.14 & 70.90 & 64.08 & 58.58 \\
NCI109          & 68.60 & \win 73.76 & 69.48 & 70.75 & \win 73.96 & 70.67 & 64.28 & 58.76 \\
ENZYMES         & 92.51 & \win 95.20 & \win 95.70 & 94.03 & 90.77 & 90.10 & 53.93 & 87.38 \\
\midrule
COLLAB          & 59.82 & 65.85 & 69.74 & 69.01 & 70.35 & \win 71.89 & 55.18 & 54.43 \\
IMDB-B          & 67.18 & 70.58 & 69.22 & \win 75.26 & \win 75.54 & 74.13 & 56.23 & 54.44 \\
IMDB-M          & 74.45 & 75.51 & 75.54 & \win 77.99 & 78.68 & 76.97 & 56.31 & 48.06 \\
\bottomrule
\end{tabularx}
}
\end{center}
\caption{ROC AUC \pk{in detecting whether a graph is real.}}\label{tbl:real-or-not}
\vspace{-6mm}
\end{table*} %
\begin{table*}[ht]
\begin{center}
{
\newcolumntype{C}{>{\centering\arraybackslash}X}
\begin{tabularx}{\textwidth}{XCCCCCCCC} %
\emph{dataset}  & $h(G)$ & $h(G)/h(\bar{K})$ & $h(G)/h(K)$ & $w(G)$ & $w(G)/w(\bar{K})$ & $w(G)/w(K)$  & \fgsd{} & \netsimile{} \\ 
\midrule
MUTAG           & \win 86.47 & 85.32 & 84.66 & 83.35 & 81.72 & 82.22 & 84.90 & 84.09 \\
PTC             & 55.30 & 52.76 & 51.16 & 54.97 & 54.53 & 53.40 & 60.28 & \win 61.26 \\
PROTEINS        & 64.89 & 65.73 & 65.36 & \win 66.80 & 65.58 & 62.27 & 65.30 & 62.45 \\
NCI1            & 66.49 & 67.44 & 64.82 & 70.78 & 67.67 & 62.19 & \win 75.77 & 66.56 \\
NCI109          & 65.89 & 66.93 & 64.78 & 69.32 & 67.08 & 63.53 & \win 74.59 & 65.72 \\
ENZYMES         & 31.99 & 33.31 & 37.19 & 40.41 & 35.78 & 28.75 & 41.58 & 33.23 \\
D\&D            & 69.86 & 68.38 & 67.09 & 68.77 & 65.39 & 65.12 & 70.47 & 64.89 \\
\midrule
COLLAB          & 68.00 & 69.42 & 69.70 & 75.77 & \win 77.24 & 67.37 & 73.96 & 73.10 \\
IMDB-B          & 68.04 & \win 70.17 & 69.45 & 68.63 & 69.33 & 61.67 & 69.54 & 69.20 \\
IMDB-M          & 40.51 & 40.34 & 40.10 & \win 42.66 & 42.00 & 39.71 & 41.14 & 40.97 \\
REDDIT-M        & \win 43.12 & 40.62 & 39.08 & 41.49 & 38.65 & 41.24 & 41.61 & 41.32 \\
REDDIT-S        & 83.67 & 81.77 & 83.73 & 84.49 & 70.47 & 79.46 & 88.95 & \win 89.65 \\
\bottomrule
\end{tabularx}
}
\end{center}
\caption{Accuracy in 1-NN Classification.}\label{tab:classification-accuracy}
\vspace{-6mm}
\end{table*} %
\subsection{Approximation quality}\label{ssec:approximation-quality}

\pk{First}, we study the quality of the approximation technique of Section~\ref{ssec:computation}.
Figure~\ref{fig:approximation-accuracy-diff} reports relative error results in terms of deviation from \pk{the exact} version, averaged over 2085 REDDIT-X graphs.
Specifically, Figure~\ref{fig:approximation-accuracy-diff}a shows the quality of the approximation varying the number of eigenvalues $k$, \pk{using} only the $k$-smallest (blue line) and the $k/2$-smallest and $k/2$-largest eigenvalues (green line). Using eigenvalues from the two \pk{ends} of the spectrum achieves consistently better performance. Figure~\ref{fig:approximation-accuracy-diff}b shows the impact of $t$ in the approximation. 
\pk{The prediction is easier at large $t$, as the spectrum converges to the constant value 1; at medium $t$ values, the approximation is harder. Still, the use of the lowest and largest eigenvalues delivers almost an order of magnitude higher accuracy than using only one side of the spectrum, vindicating our choice of approximation.}

\subsection{\pk{Identifying} real-world networks}

\pk{We devise a binary classification task of detecting whether a graph is real or synthetic.} Such tasks are \pk{critical in} anomaly detection and detecting bots and trolls in social networks.
\pk{To render the task challenging enough, rather than generating purely random graphs (which are easy to detect), we produce synthetic graphs by rewiring real ones while preserving their degree distribution, using 10 iterations of shuffling all their edges via Metropolis-Hastings sampling~\cite{peixoto2014}.
We consider a label indicating whether the graph has been rewired.
We label 80\% of the dataset and test 20\%, predicting a graph's label as that of its Nearest Neighbor (1-NN) by each similarity measure, and report the average across 100 trials. Table~\ref{tbl:real-or-not} shows our results, in terms of ROC AUC, for all datasets and measures. These results confirm the effectiveness of \thiswork.} %

\begin{table}[h!]
\vspace{-2mm}
\begin{center}
{
\newcolumntype{C}{>{\centering\arraybackslash}X}
\begin{tabularx}{\columnwidth}{XCCC} %
\multicolumn{1}{C}{} & \multicolumn{3}{c}{$k$} \\
\cmidrule(lr){2-4}
Method              & 100 & 200 & 300  \\ 
\midrule
$h(G)$              & 68.91 & 68.89 & 68.01 \\
$h(G)/h(\bar{K})$   & 62.69 & 61.74 & 61.88 \\
$h(G)/h(K)$         & 70.11 & 69.40 & \win  70.88 \\
\hline
$w(G)$              & 71.27 & 69.93 & 68.93 \\
$w(G)/w(\bar{K})$   & 64.79 & 65.81 & 65.90 \\
$w(G)/w(K)$         & 64.51 & 69.49 & \win  72.64 \\
\bottomrule
\end{tabularx}
}
\end{center}
\caption{Accuracy in 1-NN Classification with REDDIT-L.
}\label{tab:classification-largescale}
\vspace{-8mm}
\end{table} %
\subsection{Graph classification}

We now assess \thiswork{} \pk{on a traditional task of 1-NN graph classification, using labels as provided in the datasets in question and splitting training and testing as in the previous experiments.} Table~\ref{tab:classification-accuracy} reports the quality in terms of ROC AUC, averaged over 1000 trials.
\pk{Again, \thiswork is on a par with other methods.}

We additionally report in Table~\ref{tab:classification-largescale} the quality results for the largest dataset at our disposal (REDDIT-L). 
\netsimile and \fgsd cannot scale to such large graphs and therefore we do not report on them; on the other hand, \thiswork processes graphs with millions of nodes and attains good overall quality. 
\subsection{Discerning community structures}

Communities are set of vertices sharing common characteristics in terms of connectivity and attributes. Community detection~\cite{kloster2014,shi2000normalized} is one of the prototypical, yet only partially solved, tasks in graphs. An expressive graph distance should \pk{discriminate graphs with community structure from those without.}

\pk{To evaluate the expressiveness of \thiswork in terms of communities, we devise a graph classification experiment:} We generate some graphs with community structure and some without, and the classifier's task is to predict whether test set graphs have community structure. We employ a simple 1-NN classifier, as our goal is to test the representation's expressiveness rather than the performance of the classifier. %
We generate $1000$ random graphs with Poisson degree distribution $\mathcal{P}(\lambda)$ with mean degree $\lambda=10$ and fixed size. Then, we sample another $1000$ graphs from the stochastic block model (SBM)~\cite{karrer2011} with $10$ communities, following the same degree distribution. 
The stochastic block model produces graphs with clear community structure as opposed to random ones. We use 80\% of the dataset for training, and 20\% for testing, repeat the experiment with different training and testing sets, and report the average across trials. Table~\ref{tbl:comm-same-nv} reports the average quality for discerning SBM graphs by 1-NN classification in terms of ROC-AUC\@, as we vary the graph size in $(64,128,256,512,1024)$. \thiswork significantly outperforms the competitors and improves in quality on larger graphs. \pk{While discriminating very small communities is intuitively harder than distinguishing large ones, the performance of both \fgsd and \netsimile drops with increasing size, suggesting that these methods only capture local, small-scale variations.} This result verifies the capacity of \thiswork to capture global and local characteristics.
\begin{table}[!h]
\vspace{-3mm}
\begin{center}
{
\newcolumntype{C}{>{\centering\arraybackslash}X}
\begin{tabularx}{\columnwidth}{p{1.5cm}CCCCC} %
\multicolumn{1}{C}{} & \multicolumn{5}{c}{\textbf{Number of nodes} $n$} \\
\cmidrule(lr){2-6}
\emph{Method}           & 64    & 128   & 256   & 512   & 1024  \\ 
\midrule
$h(G)$                  & 57.40 & \win 68.37 & \win 77.42 & \win 82.83 & \win 84.63 \\
$h(G)/h(\bar{K})$       & 57.42 & \win 68.40 & \win 77.41 & \win 82.84 & \win 84.63 \\
$h(G)/h(K)$             & 56.71 & 67.96 & \win 77.50 & \win 83.31 & \win 85.12 \\
\midrule
$w(G)$                  & 57.47 & 66.97 & 73.95 & 78.43 & 80.26 \\
$w(G)/w(\bar{K})$       & 57.44 & 66.98 & 73.96 & 78.43 & 80.25 \\
$w(G)/w(K)$             & 56.75 & 66.26 & 73.06 & 77.05 & 78.76 \\
\midrule
\fgsd{}                 & 58.00 & 55.73 & 55.46 & 53.43 & 51.57 \\
\netsimile{}            & \win 65.73 & 61.31 & 61.51 & 61.86 & 61.58 \\
\bottomrule
\end{tabularx}
}
\end{center}
\caption{Accuracy \pk{in detecting graphs with communities.
}}\label{tbl:comm-same-nv}
\vspace{-7mm}
\end{table} 
In the second experiment we \pk{evaluate} the size-invariance of \thiswork: The task is to discriminate \pk{fixed-size} communities in graphs \pk{of increasing size}. Therefore, we sample the number of nodes $n$ from a Poisson distribution $\mathcal{P}(\lambda = n)$ with variance $\lambda$, and then generate a network with $n$ nodes. \pk{Again,} we repeat the process $1000$ times with purely random generated graphs and $1000$ with random graphs with community structure\pk{, and perform a classification experiment as before.} Table~\ref{tbl:comm-diff-nv} reports \pk{the results, for increasing~$\lambda$.}
Once \pk{again}, \thiswork outperforms the competitors. This result confirms that the normalization proposed in Section~\ref{ssec:properties} is effective in detecting community structures in graphs \pk{of different size}. %
\pk{Table~\ref{tbl:comm-diff-nv} also indicates that, while both normalizations are similarly effective with the wave kernel, the complete-graph normalization produces worse, but still competitive, results with the heat kernel.}
\begin{table}[h!]
\begin{center}
{
\newcolumntype{C}{>{\centering\arraybackslash}X}
\begin{tabularx}{\columnwidth}{p{1.5cm}CCCCC} %
\multicolumn{1}{C}{} & \multicolumn{5}{c}{$n \sim \mathcal{P}(\lambda)$} \\
\cmidrule(lr){2-6}
\emph{Method}           & 64    & 128   & 256   & 512   & 1024  \\ 
\midrule
$h(G)$                  & 54.39 & 59.01 & 60.82 & 57.99 & 53.80 \\
$h(G)/h(\bar{K})$       & 54.53 & 62.27 & 70.83 & \win 76.45 & \win 78.40 \\
$h(G)/h(K)$             & 54.37 & 60.93 & 66.86 & 68.24 & 65.23 \\
\midrule
$w(G)$                  & 56.23 & \win 63.77 & 69.57 & 71.66 & 70.34 \\
$w(G)/w(\bar{K})$       & 55.51 & \win 63.85 & \win 72.12 & \win 77.59 & \win 79.39 \\
$w(G)/w(K)$             & 56.69 & \win 64.92 & \win 71.81 & 75.91 & 77.50 \\
\midrule
\fgsd{}                 & 55.44 & 54.99 & 53.86 & 52.74 & 50.92 \\
\netsimile{}            & \win 59.55 & 56.57 & 59.41 & 66.23 & 60.58 \\
\bottomrule
\end{tabularx}
}
\end{center}
\caption{\pk{Accuracy in detecting graphs with communities, Poisson distribution of graph size.}}\label{tbl:comm-diff-nv}
\vspace{-8mm}
\end{table} 
\pk{Last, we assess the size-invariance of \thiswork in a tough regime: discriminating communities in graphs with a number of nodes {\em chosen uniformly at random}. We sample the number of nodes $n$ from the uniform distribution $\mathcal{U}(10, \lambda)$; other experimental settings remain the same. Table~\ref{tbl:comm-uniform-nv} reports the results with growing~$\lambda$. \thiswork outperforms the competitors once again, yet the results suggest that this task is more challenging. Complete\hyp{}graph normalization for the wave kernel and empty\hyp{}graph normalization for the heat kernel perform best. We conclude that, for fully size-agnostic comparisons, normalization should be chosen carefully.}

\begin{table}[ht]
\vspace{-2mm}
\begin{center}
{
\newcolumntype{C}{>{\centering\arraybackslash}X}
\begin{tabularx}{\columnwidth}{p{1.5cm}CCCCC} %
\multicolumn{1}{C}{} & \multicolumn{5}{c}{$n \sim \mathcal{U}(10, \lambda)$} \\
\cmidrule(lr){2-6}
\emph{Method}           & 64    & 128   & 256   & 512   & 1024  \\ 
\midrule
$h(G)$                  & 51.16 & 51.98 & 51.62 & 51.05 & 50.08 \\
$h(G)/h(\bar{K})$       & 51.19 & 53.36 & 56.63 & 59.10 & 59.71 \\
$h(G)/h(K)$             & 51.69 & 53.67 & 55.43 & 55.29 & 53.43 \\
\midrule
$w(G)$                  & 52.38 & 55.89 & 59.81 & 61.18 & 59.13 \\
$w(G)/w(\bar{K})$       & 51.61 & 54.67 & 57.83 & 60.06 & 61.01 \\
$w(G)/w(K)$             & 52.63 & 57.48 & \win 62.85 & \win 67.95 & \win 71.19 \\
\midrule
\fgsd{}                 & 57.92 & 55.62 & 54.94 & 52.74 & 52.15 \\
\netsimile{}            & \win 63.63 & \win 58.31 & 55.75 & 54.34 & 53.11 \\
\bottomrule
\end{tabularx}
}
\end{center}
\caption{\pk{Accuracy in detecting} graphs with communities, Uniform distribution of graph size.}\label{tbl:comm-uniform-nv}
\vspace{-7mm}
\end{table} 

\begin{figure*}[!t]
\centering
\begin{tabular}{c@{\hskip 1in}c}
\subfloat[Protein network. $|V|=190, |E|=744, \diam=9$.]{
\includegraphics[width=0.7\columnwidth]{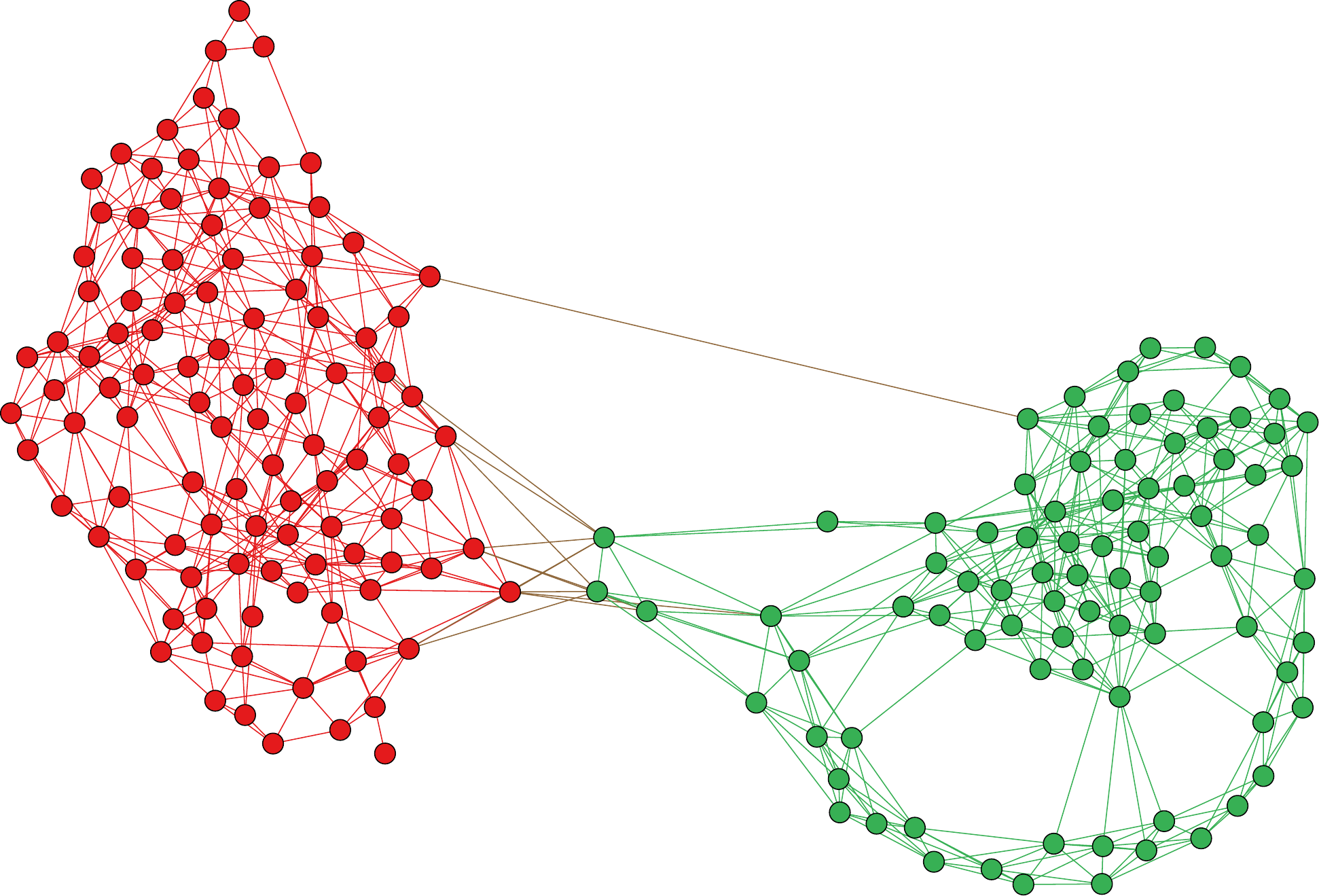}\label{sfig:protein}} & %
\subfloat[Enzyme. $|V|=125, |E|=141, \diam=32$.]{
\includegraphics[width=0.6\columnwidth]{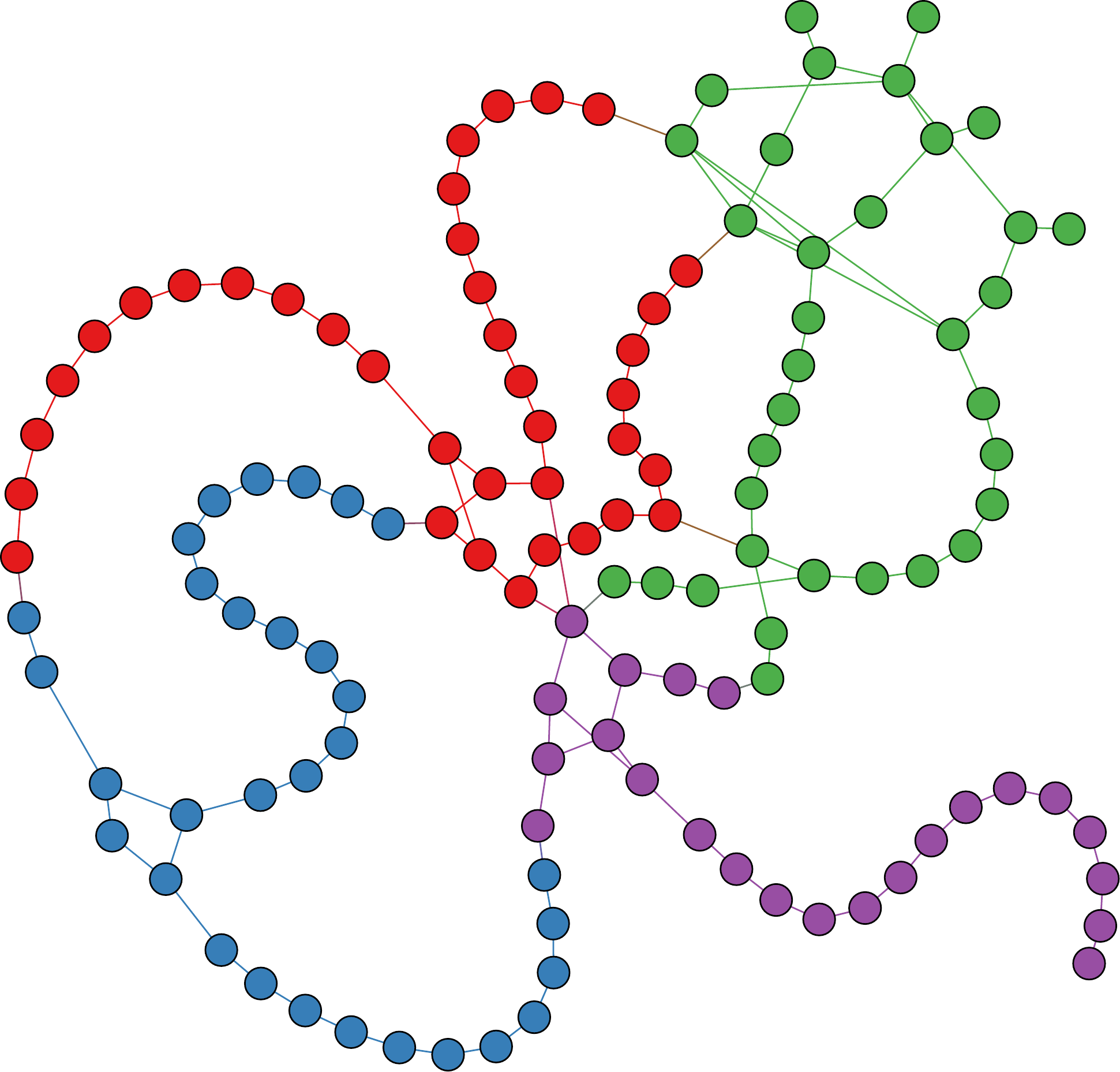}\label{sfig:enzyme}}
\end{tabular}
\vspace{-2mm}
\caption{Two most dissimilar graphs by $h(G)/h(\bar{K})$ across all small-size graphs in datasets used. Communities are colored.}\label{fig:casestudy}
\vspace{-2mm}
\end{figure*}

\subsection{Case Study}

\pk{Here, we visualize the discovery potential of using \thiswork as a similarity measure. We run a {\em furthest pair query\/} on all graphs in our collection bar those in the REDDIT-L dataset, which are harder to visualize readably; thereby, we discover the two graphs of {\em lowest similarity}, using the normalized heat kernel.} Figure~\ref{fig:casestudy} shows those two graphs: a {\em protein interaction network\/} from the D\&D dataset and an {\em enzyme's tertiary structure\/} from the ENZYMES dataset. 
These two graphs are \pk{conspicuously} different across multiple scales, from local patterns to global structure. 
The interaction network in Figure~\ref{sfig:protein} is a small-world graph with large clustering coefficient ($0.425$) and small diameter ($\diam = 9$), whereas the tertiary structure in Figure~\ref{sfig:enzyme} is a connected collection of paths of big length with negligible clustering coefficient ($0.006$) and large ($\diam=32$) diameter.

\begin{figure}[!t]
\centering
\resizebox{0.7\columnwidth}{!}{%
\begin{tikzpicture}
    \begin{axis}[
    ymode=log,
    xmode=log,
    ylabel=time (s),
    xlabel=$n$,
]
\addplot[ultra thick,color=cycle1] coordinates {
(5000, 7.291010149)
(10000, 12.30674258)
(15000, 18.62399152)
(30000, 30.93526703)
(50000, 50.4202837)
(100000, 118.3770651)
(150000, 177.5655977)
(300000, 355.1311953)
(500000, 598.8853255)
(1000000, 1181.770651)
};
\end{axis}
\end{tikzpicture} %
}%
\vspace*{-3mm}
\caption{\pk{Time to compute} $300$ eigenvalues on both ends of the spectrum and the approximation of $h(G)$.}\label{fig:scalability}
\vspace{-4mm}
\end{figure}
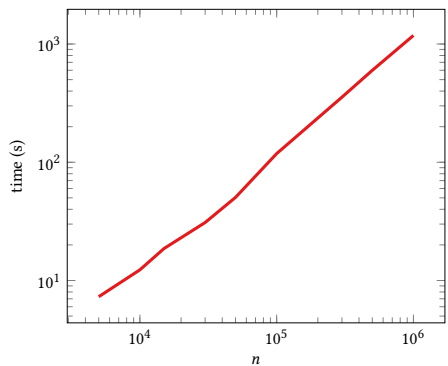

\subsection{Scalability}\label{ssec:scalability}

Last, we corroborate the scalability of \thiswork. \pk{Figure~\ref{fig:scalability} shows} the time to compute $300$ eigenvalues for the approximation of $h(\graph)$ on graphs of increasing size from the REDDIT dataset. Our method computes the similarity on graphs of one million ($10^6$) nodes in only $16$ minutes, while previous methods could not complete the process within one day. This result \pk{illustrates the fitness of} \thiswork a scalable comparison method among real-size graphs. %
\section{Conclusions}

We proposed \thiswork, a representation for graph comparison relying neither on (i) graph alignment operations, nor on (ii) computationally\hyp{}demanding kernel computations, nor on (iii) supervised representation learning.
\pk{\thiswork{} is a multi-scale heat trace signature of a graph Laplacian spectrum,
which lower-bounds the Gromov\hyp{}Wasserstein distance as it incorporates heat traces covering all scales.}
We derived a novel approximation of \pk{heat traces}, rendering \thiswork efficiently computable, and a normalization scheme, rendering it size\hyp{}invariant.
To our knowledge, this is the first graph \pk{representation} that achieves these properties and allows \pk{for} comparisons at multiple scales.
Our experiments show that \thiswork outperforms \netsimile{} and \fgsd{}, two state\hyp{}of\hyp{}the\hyp{}art representation-based methods for graph comparison, on a variety of graph collections on community detection and graph classification.

\balance
\bibliographystyle{ACM-Reference-Format}
\bibliography{bibliography} 
\end{document}